\documentclass{aa}  
\usepackage{graphicx}
\usepackage{txfonts}
\usepackage{newtxmath}
\usepackage[hidelinks,colorlinks=true,linkcolor=blue,citecolor=blue]{hyperref}
\newcommand{\vect}[1]{\boldsymbol{#1}}
\defcitealias{gehlot2020}{G20}
\newcommand{\lroman}[1]{\romannumeral#1\relax}
\begin{document}

   \title{Transient RFI environment of LOFAR-LBA at 72-75 MHz}
   \subtitle{Impact on ultra-widefield AARTFAAC Cosmic Explorer observations of the redshifted 21-cm signal}
     
   \author{B.~K.~Gehlot
          \inst{1}\fnmsep\thanks{Email: kbharatgehlot@gmail.com},
          L.~V.~E.~Koopmans\inst{1},
          S.~A.~Brackenhoff\inst{1},
          E.~Ceccotti\inst{1},
          S.~Ghosh\inst{1},
          C.~H\"{o}fer\inst{1},
          F.~G.~Mertens\inst{2}, 
          M.~Mevius\inst{3}, 
          S.~Munshi\inst{1}
          A.~R.~Offringa\inst{3,1},
          V.~N.~Pandey\inst{3,1},
          A.~Rowlinson\inst{4,3},
          A.~Shulevski\inst{3,1}, 
          R.~A.~M.~J.~Wijers\inst{4}, 
          S.~Yatawatta\inst{3},
          \and
          S.~Zaroubi\inst{5,1}.
          }

   \institute{Kapteyn Astronomical Institute, University of Groningen, PO Box 800, 9700AV Groningen, The Netherlands.
      \and
      LERMA, Observatoire de Paris, PSL Research University, CNRS, Sorbonne Universit\'e, F-75014 Paris, France.
      \and
      ASTRON, Netherlands Institute for Radio Astronomy, Oude Hoogeveensedijk 4, 7991 PD, Dwingeloo, The Netherlands.
      \and
      Anton Pannekoek Institute, University of Amsterdam, Postbus 94249 1090 GE, Amsterdam, The Netherlands.
      \and
      Department of Natural Sciences, The Open University of Israel, 1 University Road, PO Box 808, Ra'anana 4353701, Israel.
    }
    \titlerunning{Impact of transient RFI}
    \authorrunning{B.~K.~Gehlot et al.}
   \date{Received ---; accepted ---}

 
  \abstract
   {Measurement of the highly redshifted and faint 21-cm signal of neutral hydrogen from the Cosmic Dawn and Epoch of Reionisation promises to unveil a wealth of information about the astrophysical processes that governed the structure formation and evolution of the universe during the first billion years of its evolution.}
   {The AARTFAAC Cosmic Explorer (ACE) program utilises the AARTFAAC wide-field imager of LOFAR to measure the power spectrum of the intensity fluctuations of the redshifted 21-cm signal from the Cosmic Dawn at $z\sim18$ corresponding to the global 21-cm absorption feature reported by the EDGES experiment. Radio frequency interference (RFI) from various sources, such as aeroplane communication, contaminates the observed data and it is crucial to exclude the RFI-affected data in the analysis for any reliable detection. In this work, we solely focus on investigating the impact of non-ground-based transient RFI on the analysis of ACE observations.}
   {We use cross-power spectra and cross-coherence metrics to assess the correlation of RFI over time and investigate the level of impact of transient RFI on the 21-cm signal power spectrum estimation.}
   {We detected moving sky-based transient RFI sources that cross the field of view within a few minutes and appear to be mainly from aeroplane communication beacons at the location of the LOFAR core in the 72-75~MHz band (a part of the EDGES absorption trough), by inspecting filtered images. We find that this transient RFI is mostly uncorrelated over time and is only expected to dominate over the thermal noise for an extremely deep integration time of 3000 hours or more with a hypothetical instrument that is sky temperature dominated at 75~MHz. We find no visible correlation over different $k$-modes in Fourier space in the presence of noise for realistic thermal noise scenarios.}
   {We conclude that the sky-based transient RFI from aeroplanes, satellites and meteorites at present does not pose a significant concern for the ACE analyses at the current level of sensitivity and after integrating over the available $\sim500$ hours of observed data. However, it is crucial to mitigate or filter such transient RFI for more sensitive experiments aiming for significantly deeper integration.}


   \maketitle

\section{Introduction}

Observations of the redshifted 21-cm signal of neutral hydrogen from the Dark Ages (DA), Cosmic Dawn (CD), and Epoch of Reionisation (EoR) are expected to fill the gaps in our understanding of the cosmic structure formation \citep{furlanetto2006,pritchard2012,zaroubi2013}. Several operational (and upcoming) instruments and associated experiments, such as LOFAR (and AARTFAAC) \citep{vanhaarlem2013,prasad2016}, MWA \citep{bowman2013}, HERA \citep{deboer2017}, NenuFAR \citep{zarka2012,mertens2021},  SKA \citep{koopmans2015}, LWA \citep{eastwood2019,dilullo2020}, EDGES \citep{bowman2018}, SARAS \citep{singh2017}, LEDA \citep{bernardi2016}, REACH \citep{deleraacedo2019}, are seeking to statistically measure global and spatial properties of the faint redshifted 21-cm signal from CD and EoR. The final frontier is to measure the 21-cm signal from DA, requiring space-based instruments and instruments such as the planned FARSIDE \citep{burns2021}, LuSEE \citep{bale2023}, DAPPER\footnote{\url{https://www.colorado.edu/project/dark-ages-polarimeter-pathfinder/}}, NCLE \citep{vecchio2021}, and the ESA-led ALO missions. 

However, successfully measuring this highly redshifted and faint 21-cm signal is an extremely challenging task. It requires ultra-precise calibration of the instrument and filtering of bright contaminations with a $10^5$:$1$ dynamic range or higher. These contaminations include \lroman{1}) spectrally smooth astrophysical foregrounds that are several orders of magnitude brighter than the faint 21-cm signal \citep{bernardi2009,bernardi2010,ghosh2012}; \lroman{2}) chromatic effects of the instrument (commonly referred to as mode-mixing) caused by, for instance, the primary beam, $uv$-coverage, polarisation leakage, cross-talk between receiver elements, and 
 cable reflections that modulate the smooth foregrounds and make it challenging to distinguish them from the signal of interest \citep{datta2010,morales2012,vedantham2012,trott2012,hazelton2013,dillon2014}; \lroman{3}) contamination effects in the data due to ionospheric phase errors \citep{mevius2016,jordan2017}; \lroman{4}) systematic errors due to imperfect processing of the data, for example, calibration and foreground removal \citep{barry2016,patil2016,ewall-wice2017}; and \lroman{5}) radio frequency interference (RFI) due to human activities \citep{offringa2013,offringa2015,offringa2019a,whitler2019,vruno2023}.

Radio frequency interference is a crucial challenge in 21-cm cosmology experiments and comes from sources including, but not limited to, radio broadcasts such as FM, digital TV and audio broadcasts, satellite communication (geostationary and polar), aircraft communication, interference from electricity generation and transmission architecture such as wind and solar farms, and noisy transformers. Additionally, instruments with an ultra-wide field of view (FoV) such as AARTFAAC, LWA, and MWA can be significantly more prone to RFI originating at lower elevations, that is, those on the horizon or in the sky, than instruments with a narrower FoV (for example, LOFAR, HERA, and SKA). 

We can broadly classify the RFI signatures from these sources into two categories based on their appearance in the data: (a) Stationary RFI sources: RFI signatures (whether narrowband or broadband) from a majority of sources such as FM, DTV, DAB, and geostationary satellites are usually stationary in time, frequency, and spatial location with relatively well-known frequency bands; (b) transient RFI sources: RFI signatures from polar and low-Earth orbit satellites, aircraft communication channels, drones, and reflections of radio signals off moving objects such as meteorites, etc. These may be considered transient RFI because of their spatially, temporally, and sometimes spectrally transient nature. Because of their transient nature, the impact of these RFI sources on the 21-cm signal has so far, to our knowledge, not yet been studied.

Frequency channels with stationary RFI contamination can be comparatively easily flagged during processing or avoided altogether during observations. Most RFI flagging algorithms use some form of outlier detection in post-correlation dynamic spectra and are designed to detect and flag stationary and transient (bright) RFI signatures \citep{middelberg2006,offringa2010,offringa2012,prasad2012,peck2013}. However, faint RFI below the noise level is difficult to detect, and techniques that leverage the spectral smoothness of the foregrounds and the Gaussian nature of instrumental noise are required for faint RFI rejection for a given dataset. \cite{offringa2015} shows an approach to detect and flag faint RFI using baseline-integrated statistics (for example, the standard deviation of dynamic spectra along all baselines) and performing outlier detection. \cite{wilensky2019} developed another technique, called ``SSINS'', that removes the sky component in the visibility spectra by subtracting consecutive time samples and averaging multiple baselines incoherently to lower the noise, and then performing RFI detection on the sky subtracted and averaged spectrum. Other advanced techniques for RFI mitigation include compressive statistical sensing \citep{cucho-padin2019}, mitigation using polarisation information of interference signals \citep{yatawatta2020}, machine learning techniques \citep{vos2019}, as well as pre-correlation interference mitigation techniques. Readers can refer to \cite{an2017} and \cite{baan2019}, for example, for a review of pre- and post-correlation RFI mitigation techniques.

In this work, we perform an investigative study of the (largely) sky-based transient RFI environment of the LOFAR-AARTFAAC Low-Band Antenna (LBA) system in the frequency range of 72-75~MHz and its impact on the analyses to measure the power spectrum of the redshifted 21-cm signal at redshift $z\sim18$. More specifically, we focus on the characterisation of the spatially and temporally transient RFI that might require different methods to detect and remove. We do not focus on mitigation techniques for such sky-based transient RFI and defer that to future analyses. For the RFI analysis, we use typical 15-minute datasets from eight different nights that were also used to report the first 21-cm signal power spectrum limits from the AARTFAAC Cosmic Explorer (ACE) by \cite{gehlot2020} (referred to as \citetalias{gehlot2020} hereafter). We investigate the impact of transient RFI using Fourier space statistics such as their cross-coherence, cylindrical, and spherical power spectra.  

\section{The LOFAR-AARTFAAC Cosmic Explorer Programme} \label{sec:ACE_programme}

The EDGES collaboration \citep{bowman2018} reported a putative detection around 78~MHz of a 21-cm absorption feature, during CD, in the otherwise smooth frequency spectrum of the low-frequency radio sky. Such an absorption feature can result from the first generation of stars that appeared during the CD ($z\sim18$), coupling the spin temperature of neutral hydrogen to the gas temperature. Since the latter is below the Cosmic Microwave Background (CMB) temperature, the 21-cm signal is observed in absorption against the CMB. This putative detection is unusual in the sense that it is unexpectedly deep and also wider than the most extreme scenarios predicted by standard astrophysical simulations. Hence, non-standard (``exotic'') models are required to explain the unusual nature of the reported detection. Several concerns were raised by \cite{hills2018,bradley2019} regarding the validity of the signal. Furthermore, the claimed detection was recently contested by the SARAS collaboration \citep{singh2022} whose observations are inconsistent with the EDGES global 21-cm absorption feature at a marginally significant level. However, if the EDGES detection is confirmed, it may unveil the presence of previously unknown fundamental physical processes such as the interaction between dark matter and baryonic matter, or the presence of a strong radio radiation background in addition to the CMB \citep{barkana2018,ewall-Wice2018,dowell2018,feng2018,fialkov2019,reis2021}. Additionally, a deeper global 21-cm absorption feature is expected to enhance the brightness-temperature fluctuations of the 21-cm signal from redshifts corresponding to the absorption trough \citep{bowman2018}. This makes it possible to detect these brightness temperature fluctuations in a significantly shorter integration time. To test this hypothesis, in 2018 we commenced the ACE programme using the AARTFAAC-12 wide-field imager of the LOFAR telescope to observe the radio sky (for up to a thousand hours) between 72-75~MHz, within the absorption feature. The key science goal of the ACE programme is to measure or set the first limits on the power spectrum of the redshifted 21-cm signal from the Cosmic Dawn at $z\sim18$ \citepalias{gehlot2020} and constrain/exclude non-standard models that could explain the unusual nature of the EDGES absorption feature.

\subsection{The wide-field imaging mode}

The Amsterdam Astron Radio Transients Facility and Analysis Center (AARTFAAC) is an independent system and observing mode that utilises the LOFAR in-field hardware (receivers and the signal chain) to perform full-sky imaging with the LOFAR-LBA dipoles\footnote{For the HBA observations, the FoV$\sim25^\circ$ at 150~MHz} using an independent correlator. AARTFAAC can piggyback on ongoing LOFAR observations by tapping off the voltage streams from individual LBA dipoles (or HBA tiles) before beam-forming, which are then transported to a GPU-based correlator located at the Center for Information Technology of the University of Groningen. The AARTFAAC system has two operating modes, \texttt{A6} where 288 receiver elements\footnote{The receiver element can be either LBA dipoles or HBA tiles, with different user selectable configurations. \texttt{LBA\_OUTER} is the default configuration for AARTFAAC observations.} from the six innermost stations (``superterp") of the LOFAR core are correlated, and \texttt{A12} where 576 receiver elements from the 12 innermost stations are correlated. Correlating signals from individual LBA dipoles provide an instantaneous full-sky FoV which is optimal for large angular-scale 21-cm cosmology experiments. Although AARTFAAC can observe in the full LBA frequency range of 10-90~MHz, the instantaneous bandwidth is currently limited to 3.1~MHz\footnote{The bandwidth is divided into 16~subbands of 195.3~kHz width and each subband can have a maximum of 64~channels of 3~kHz width.} due to the limited bandwidth of the signal transport chain. The correlator has an integration time of 1~second and produces XX, XY, YX, YY auto and cross-correlation products, where X and Y correspond to the orthogonal polarisations of the dual-polarisation receivers. Unlike standard tracking and beam-formed observations with LOFAR,  AARTFAAC observes the sky in the drift-scan mode and stores the correlation products as raw binary files. Readers are referred to \cite{vanhaarlem2013,prasad2016,kuiack2019,shulevski2022} for more information about the LOFAR and AARTFAAC observing modes.

\subsection{Observational setup}
For the ACE observing campaign, we observed the northern sky in drift scan mode targeting the $72.36-75.09$~MHz frequency range ($z=17.9-18.6$) by placing 14~subbands (with 3 channels of 65~kHz width each) together. We use the remaining two subbands as frequency outriggers placed about 4~MHz apart on either side of the spectral window. The outrigger subbands can be used for quantification of frequency smooth foregrounds and systematics. Due to the large number of antenna elements, the AARTFAAC data volumes in \texttt{A12} mode are enormous. Therefore, we limit ourselves to a frequency resolution of 65~kHz (3~channels per subband) and 1~second correlator integration time as a compromise between large data volume and the total number of frequency channels (42 channels for the targeted spectral window + 6 outrigger channels) that are sufficient for 21-cm power spectrum analyses. The data rate is about 0.84~TB per hour of observation with the \texttt{A12} mode, for the correlator time and frequency resolution of 1~second and 65~kHz respectively. The ACE observing campaign lasted for three LOFAR observing cycles (10, 11 and 12) and around 500~hours of data (between sunset and sunrise) were obtained. The recorded drift-scan data is split into local sidereal time (LST) ``slices'' of 15~minutes duration and each LST bin is phased to the direction that transits the zenith halfway through the selected LST slice. We refer to this observing strategy as ``semi drift-scan'' mode \citepalias{gehlot2020}. Table~\ref{tab:telescope_details} summarises the observing setup of the ACE program.

\begin{table}
	\centering
	\caption{ACE observing campaign setup details}
	\label{tab:telescope_details}
	\begin{tabular}{ll} 
		\hline		
		\textbf{Parameter} & \textbf{Value} \\	
		\hline
            Telescope & AARTFAAC (\texttt{A12} mode) \\
            Number of antennas & 576 (dual polarisation) \\
            Baseline range & $3 - 1290$~m \\
            Frequency range & $72.36 - 75.09$~MHz \\
            Outrigger frequencies & 68.4~MHz and 79.1~MHz \\
            Number of subbands & 14 (target) + 2 (outrigger) \\
            Channels per subband & 3 (Total channels: 48) \\
            Channel width & 65~KHz \\
            Correlator integration time & 1~second \\
            Observing mode & ``semi drift-scan'' \\
            Total data recorded & $\sim$500~hours \\
            Recorded data volume & $\sim$420~TB (uncompressed)\\
		\hline
	\end{tabular}
\end{table}

\section{Observations and data reduction}

In this work, we limit ourselves to eight observing blocks of 15~min duration spanning the LST range $23.5-23.75$~h. This LST range corresponds to the LST-bin1 in the analyses presented in \citetalias{gehlot2020} and these observing blocks are referred to as ``time-slices''. Each of these time-slices is processed with an updated version of the ACE pipeline as used by \citetalias{gehlot2020}. The eight time-slices are processed as follows:\\
\newline
\textit{Phase tracking:} We phased the drift-scan data to the direction ($\alpha,\delta$)$=$(23h37m30s, +52d37m21.44s) which transits the zenith halfway through the LST range $23.5-23.75$~h. The phased data was then converted to a measurement set (MS) format table and stored. Phasing of the drift-scan data and conversion to MS is done using the \textsc{aartfaac2ms}\footnote{\url{https://github.com/aroffringa/aartfaac2ms}} package \citep{offringa2015}.\\
\newline
    \textit{Initial RFI flagging and time averaging:} We used \textsc{aoflagger}\footnote{\url{https://gitlab.com/aroffringa/aoflagger}} \citep{offringa2010,offringa2012} to detect and flag the RFI affected data. We have devised a customised flagging strategy for AARTFAAC data instead of using a generic flagging strategy. The strategy is specifically tuned to detect transient RFI in noisy data (key parameters: \texttt{base\_threshold}=3.0, \texttt{iterations}=5, \texttt{transient\_threshold\_factor}=0.5). This step flags the bright RFI per baseline that appears above the noise level. After flagging, the data is averaged to a lower time resolution of 4~seconds. Any averaged time-frequency blocks with missing samples are also flagged to avoid low-level artefacts due to flagging and averaging \citep{offringa2019a}. Any bright RFI that severely impacts further processing and analysis is flagged during this step. We note that the analysis of transient RFI in this paper is therefore the RFI that is not detected during this customised per baseline visibility-based flagging step.\\ 
    \newline
    \textit{Initial calibration:} This step performs the initial direction-independent calibration of the flagged and averaged data to set the flux scale (for every frequency channel) and direction-independent phase. We used a sky model consisting of about 9990 compact sources extracted from VLSS \citep{kassim2007} and Cambridge radio catalogues \citep{laing1983} with a flux-density greater than 2~Jy at 74~MHz after extrapolating with the corresponding spectral models. The flux-density threshold we have used for the sky model is four times the classical confusion limit for the \texttt{A12} system at 74~MHz. We also added the bright ``A-team'' sources, Cas\,A, Cyg\,A, and Tau\,A\footnote{The models of these sources consist of delta functions and Gaussians to describe compact and extended emission, respectively. These models were obtained using LOFAR-LBA observations and are a part of the list of calibrator sources in the LOFAR Initial Calibration (LINC) pipeline. \url{https://git.astron.nl/RD/LINC}} (with an angular extent of several arcminutes) to the sky model in addition to the 9990 sources. We used the \texttt{gaincal} task of the \textsc{dp3} package\footnote{\url{https://git.astron.nl/RD/DP3}} \citep{vandiepen2018} to perform the calibration. We used a solution interval of 65~kHz and 4~seconds (the same as the frequency and time resolution of the datasets after flagging and averaging) and excluded baselines shorter than $20\lambda$ from the calibration. The lower baseline cut helps mitigate two issues: first, our calibration model does not include large-scale diffuse emission, which is partly polarised and dominates these short baselines, and excluding short baselines from calibration thus avoids biases due to the missing diffuse component in the sky model. Second, the excluded baselines are dominated by intra-station baseline pairs which share the same electronic cabinets and might be affected by low-level cross-talk between dipoles. We note that the expected noise level for the 65~kHz and 4~seconds solution interval is about 2500~Jy (for an LBA-dipole SEFD of $\sim1.8$~MJy), and a signal-to-noise ratio of about 36 per solution and baseline for the given calibration model. This step enables the detection of outlier antennas, baselines, and times, as discussed in the next step, and the data is re-calibrated (discussed in the following steps) after flagging these bad antennas/baselines to improve the overall quality of the calibration.\\    
    \newline
    \textit{Statistic based visibility flagging:} The \textsc{aoflagger} software also calculates various quantities: RFI percentage, mean, and standard deviation of visibilities (and their frequency difference) along different axes such as time, frequency, and baselines. These statistics were calculated using the data after the initial calibration in the previous step and can be used to identify outlier antennas, baselines, times, and frequency channels and flag them. We use a robust standard deviation estimate, $\sigma \approx 1.4826\,\text{MAD}$ (median absolute deviation), to flag outlier baselines (with $6\sigma$ threshold for RFI percentage and $5\sigma$ for the rest of the statistics) and times ($4\sigma$ threshold for all statistics).\\
    \newline
    \textit{Bandpass calibration}: After flagging visibility outliers, we performed a bandpass calibration on the uncalibrated data after flagging and averaging with additional flags from the previous step applied. We used the same sky model as in the initial calibration for the bandpass calibration but with a 15~minute solution interval instead of 4~seconds and a single gain solution was obtained per 65~kHz frequency channel. This step allows us to capture rapid direction-independent (thus independent of the 21-cm signal) spectral variations in the global bandpass (time-averaged) of the antenna elements and correct for these variations.\\   
    \newline
    \textit{Direction-independent (DI) calibration}: After correcting the visibilities for the bandpass shape, a direction-independent calibration step was performed. This step is similar to initial calibration and uses the same calibration model and settings, that is, a solution interval of 4~seconds and 65~kHz, and the same baseline cut of $20\lambda$. Repeating the DI-calibration step after flagging bad antennas improves the overall quality of the calibration. This step corrects for the direction-independent effects on both short time and frequency scales.\\   
    \newline
    \textit{Direction-dependent (DD) calibration}: The next data-processing step was to remove the dominant sources Cas\,A and Cyg\,A, each with a flux-density of about 17~kJy (at 74~MHz), from the calibrated data. We used the \texttt{DDECal}\footnote{Readers are referred to \cite{gan2023} for a description of the algorithm.} task of \textsc{dp3} to perform the direction-dependent calibration, and obtain the direction-dependent gains towards these two sources. These gains are then used to subtract the sky model components of the two sources from the calibrated visibilities. For this step, the sky model (the same as in the DI calibration) is split in 3~directions, that is, Cas\,A, Cyg\,A, and ``main'', where the main direction contains all other sources in the sky model. Only Cas\,A and Cyg\,A are subtracted from the visibilities. We did not calibrate towards the direction of Tau\,A because of the insufficient signal-to-noise. The time and frequency solution intervals were kept the same as during the DI calibration. We excluded baselines shorter than $60\lambda$ in the calibration and performed a constrained optimisation in which the direction-dependent gains are constrained to be smooth on a 3 MHz scale. We used a different baseline cut in direction-dependent calibration to avoid overlap between the baselines used for calibration and power spectrum analysis ($20-60\lambda$). These settings are used to avoid over-fitting during the direction-dependent calibration and to perform an accurate subtraction of the bright sources without affecting the rest of the emission (whether compact or diffuse) on baselines shorter than $60\lambda$ \citep{sardarabadi2019,mevius2022}. \\
    \newline
    \textit{Post-calibration outlier detection and flagging}: After subtracting Cas\,A and Cyg\,A, we ran \textsc{aoflagger} once more on the subtracted data and also repeated the statistics-based flagging (with the same settings used previously) to identify and flag the outlier visibilities, baseline pairs, and times that are badly calibrated or behave differently from the rest. After the data-processing steps, each night have typically around $20-30$~per cent flagged data in the baseline range of $20-60\lambda$, except two nights for which about $35$ and $40$ per cent of data, respectively, are flagged.\\
    \newline
    \textit{Imaging}: After the data are calibrated and visibility outliers are flagged, they are ready to be imaged. We used \textsc{wsclean}\footnote{\url{https://gitlab.com/aroffringa/wsclean}} \citep{offringa2014,offringa2017} to image the visibilities. We used the \texttt{w-gridder} algorithm \citep{arras2021,haoyang2022} implementation in \textsc{wsclean} to grid and image the data with the gridder-accuracy of $10^{-4}$. The \texttt{w-gridder} algorithm dynamically sets gridding kernel parameters (kernel size, oversampling, number of w-layers) for a given gridder-accuracy to achieve optimal and high-fidelity imaging. We do not perform any deconvolution on the images. We used only the $20-60\lambda$ baselines for imaging because the shortest baselines may be affected by low-level cross-talk and the $uv$-sampling becomes relatively sparse for baselines longer than $60\lambda$ (about 250~meters). The longer baselines are also used during the DD-calibration step, making the excluded baselines non-optimal for 21-cm cosmology analyses. Every 1-minute in the LST interval and 65~kHz channel was imaged separately such that 15 image cubes (with 42 channels each) were produced for each of the eight time-slices. The key imaging parameters are listed in Table~\ref{tab:imaging_details}.

\begin{table}
	\centering
	\caption{Imaging parameters for \textsc{wsclean}}
	\label{tab:imaging_details}
	\begin{tabular}{ll} 
		\hline		
		\textbf{Parameter} & \textbf{value} \\	
		\hline
		Baselines & $20-60\lambda$ \\
		Pixel size & 12~arcmin\\ 		
		Number of pixels & $576\times 576$\\
		Gridder & w-gridder \\
            Gridding kernel & modified ES kernel \\
            Gridder accuracy & $10^{-4}$ (default)\\
		Weighting & Natural \\
            $uv$-pixel size & $0.5\lambda\times0.5\lambda$ \\
		\hline
	\end{tabular}
\end{table}

\begin{figure*}
    \centering
   \includegraphics[width=1\textwidth]{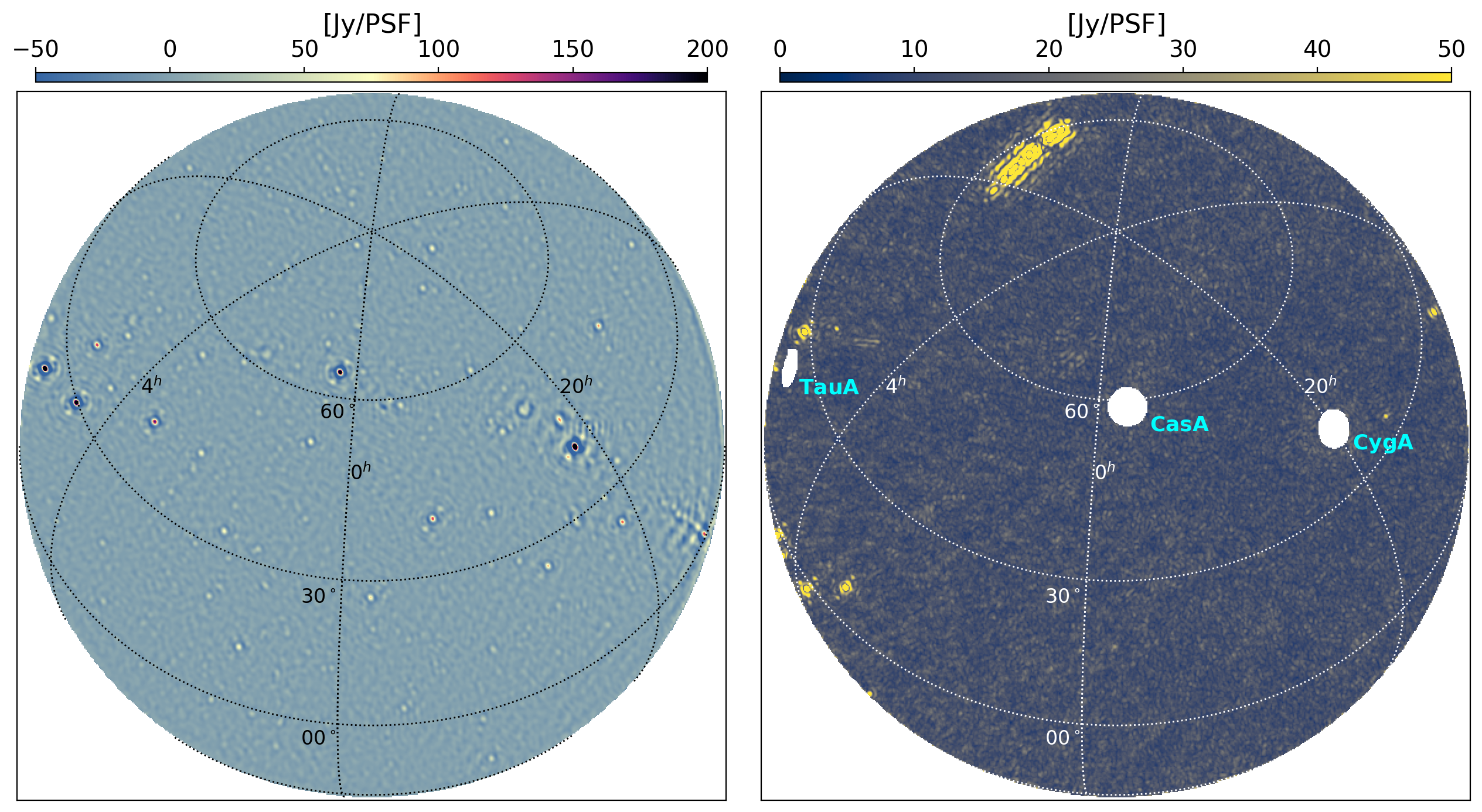}
    \caption{Images produced with a subset of ACE data. The Left panel shows a dirty continuum image of the sky produced with a one-minute subset of data from one of the nights used in the analysis. The colour scale of the image has been saturated to highlight a wide range of fluxes. The Right panel shows a spatial map highlighting the transient RFI over the one-minute duration, extracted from the same subset of data used to produce the continuum image in the left panel. Blank pixels in the panel correspond to the bright A-team sources, Cas\,A, Cyg\,A, and Tau\,A. Additionally, the $5^\circ$ region above the horizon has been excluded from both images. The bright pixels are narrow-band emissions deviating from the continuum emission shown in the left panel and are likely to be from the transient RFI. The stretched structure in the top part of the spatial map is likely an aeroplane and moves slowly over the sky (several minutes across the field).}
    \label{fig:spatial_map}
\end{figure*}

\section{Detection of transient RFI}\label{sec:rfi-extraction}

\begin{figure*}
    \centering
   \includegraphics[width=\textwidth]{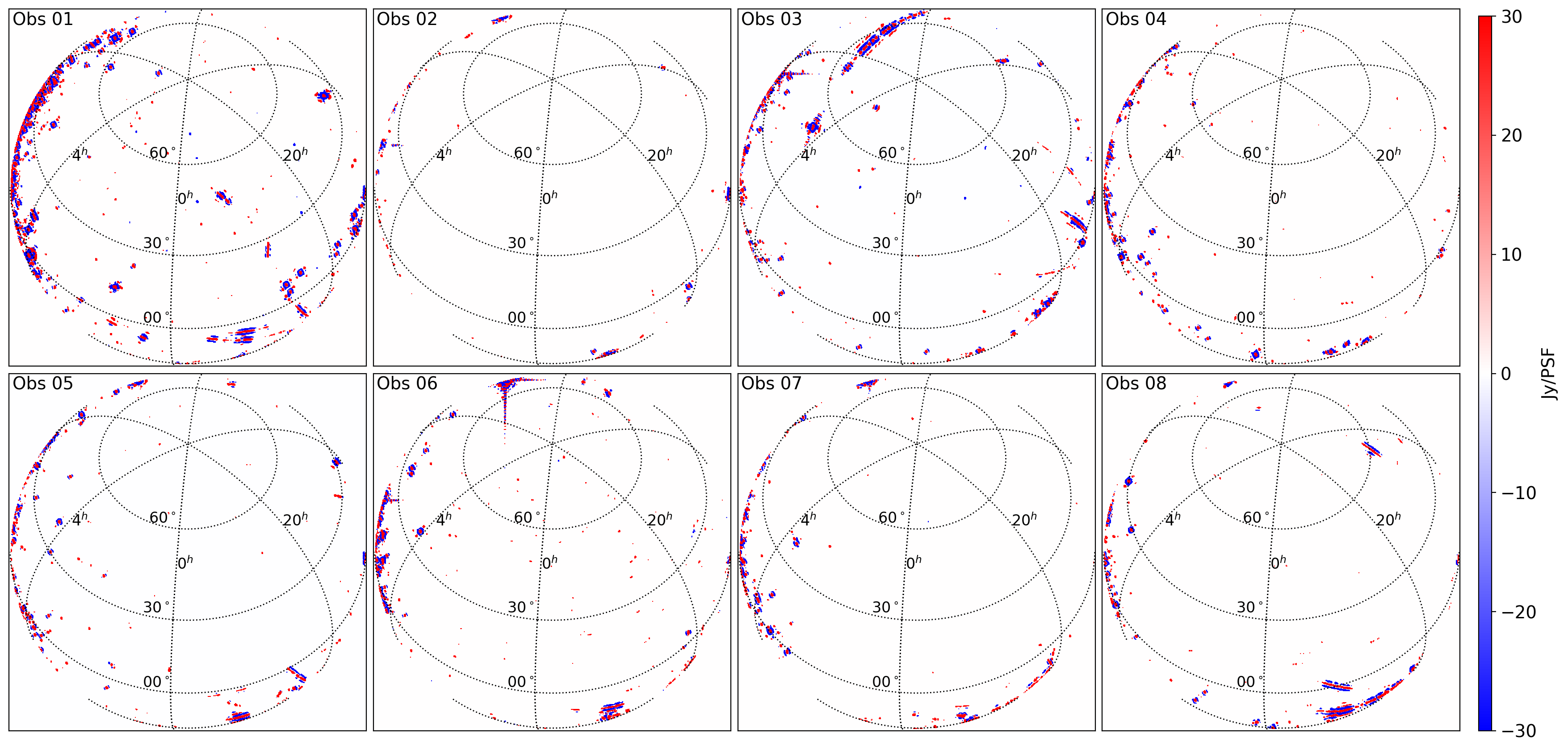}
    \caption{Stokes $I$ images of the transient RFI detected using the method described in Sect.~\ref{sec:rfi-extraction}. For visualisation purposes, RFI from each channel and time step (1~minute) is plotted on the same panel for every night. Each panel corresponds to a different night used in the analysis. Similar to Fig~\ref{fig:spatial_map}, the $5^\circ$ region above the horizon has been excluded from the images.}
    \label{fig:RFI_per_night}
\end{figure*}

Although our flagging strategy is very effective in removing RFI that is stationary and above the thermal noise, the impact of transient RFI has never been studied in detail to the best of our knowledge. To investigate the impact of transient RFI on 21-cm cosmology analyses, the RFI in each spectral image cube needs to be highlighted and extracted. To achieve this, we produce a 3D mask for the pixels that are dominated by spectrally smooth emission associated with foregrounds, sidelobe structure due to instrumental chromaticity, and spectrally uncorrelated noise. This 3D mask is then applied to the corresponding image cube to highlight the transient RFI and mask the pixels dominated by the foregrounds and the noise. We follow the steps listed below to produce the mask and residual cubes:
    \\
    \newline
    First, we subtracted the continuum emission (multi-frequency synthesis or MFS image produced with all channels combined) in a given 1-minute cadence image cube. Since the \texttt{A12} system has a uniform and dense uv-sampling in the $20-60\lambda$ baseline range, the PSF does not change significantly over the 2.7~MHz frequency band. Thus, the residual cube, to the first order, contains any emission significantly deviating from the pixel-average continuum emission, frequency-dependent PSF sidelobes, and noise. This emission is expected to be dominated by non-stationary sources. 
    \\
    \newline
    For every line of sight in the residual image cube, we selected voxels with the highest ($F_{\text{max}}$) and the lowest flux density ($F_{\text{min}}$) along the frequency direction and added them in quadrature, that is,  $F_{\text{total}}=\sqrt{0.5(F_{\text{max}}^2 +F_{\text{min}}^2)}$, to obtain a spatial map of outliers. This method highlights the bright positive and negative outliers without enhancing the noise-dominated voxels. Based on qualitative inspection of RFI spatial maps, we found that this metric performs better in terms of separating residual foregrounds and bright (narrow-band) outliers compared to the traditional sigma thresholding method using standard deviation which seems to falsely detect foregrounds as transient RFI. We excluded the directions more than $85^\circ$ away from the phase centre, to avoid false positives due to projection-based effects and proximity to the horizon in the mask creation process. We note that this mask excludes transient RFI from very close to the horizon from the analysis. We also masked the $3^\circ$ region around the bright sources Cas\,A, Cyg\,A, and Tau\,A to avoid any false detection of RFI, due to calibration and imaging artefacts. An example of a spatial map is shown in Fig.~\ref{fig:spatial_map}. In essence, this map highlights the combined level of deviations of the highest peaks and deepest valleys along the frequency axis for every line of sight. We have also provided movies of the spatial maps looping over 1-minute duration timesteps, for all eight LST time-slices used in the analysis as supplementary material.
    \\
    \newline
    Next, we used the spatial map produced in the previous step and selected the pixels that are $12\sigma$ (robust standard deviation estimator based on median absolute deviation) higher than the median of all pixels of the spatial map. The selected pixels correspond to the location of the RFI source(s) in the sky. For every such pixel, we performed an outlier detection with more than $6\sigma$ deviation from the median along the frequency direction and masked all pixels with a non-detection (including the region $<5^\circ$ above the horizon). This provides a 3D mask for a given image cube where everything is masked except for the transient, narrow-band RFI peaks and tracks localised in image space. We note that this step only selects narrow-band transient RFI that remains undetected by any RFI detection and flagging steps. Although the RFI, irrespective of its nature, could impact the 21-cm power spectrum measurement, we expect the bright RFI (whether narrow-band or broadband) also to affect the calibration process, resulting in bad calibration. Such badly calibrated data is likely to be rejected by flagging steps that are performed post-calibration. 
    \\
    \newline
    We used the final 3D mask to set all the masked pixels that were dominated by the sky emission and noise to zero in the spectral image cube, leaving only the transient RFI structures. The RFI detection process was performed on Stokes $I$ image cubes at a 1~minute time cadence for all 8 observations.\\
 \newline
 Figure~\ref{fig:RFI_per_night} shows Stokes~$I$ images of extracted RFI for the observations corresponding to the eight nights. For the sake of visualisation, we have added the RFI images for all timesteps and frequency channels to produce a combined Stokes~$I$ RFI map for each night. For a single timestep, the majority of the transient RFI peaks that we observe have a flux density between 10-100~Jy/PSF. The filtered Stokes~$I$ RFI cubes are subsequently used to investigate the impact of this transient RFI on the ultra-widefield 21-cm cosmology experiment observations as described in Sect.~\ref{sec:ACE_programme}.

\section{Impact of transient RFI}

In this section, we discuss the impact of the transient RFI that we observed in the data on 21-cm power spectrum analysis. We use the \texttt{ps\_eor}\footnote{\url{https://gitlab.com/flomertens/ps_eor}} package to estimate the power spectra and cross-coherence of the RFI image cubes produced in the previous section. This package is extensively used for LOFAR-EoR \citep{mertens2020} and ACE power spectrum analyses \citepalias{gehlot2020}. Power spectrum calculations involve the following steps: trimming image cubes to a user-provided FoV, applying a spatial taper to the trimmed image cubes, converting image flux to temperature units, spatial Fourier transforming the trimmed and tapered image cubes to obtain gridded visibility cubes,  Fourier transforming the gridded visibilities along frequency, and averaging the data in Fourier space to obtain cylindrical/spherical power spectra. We follow the above-mentioned process for most of the analysis unless specified otherwise. 

\subsection{RFI correlation over time}\label{subsec:rfi_correlation_time}
The RFI contamination, regardless of its properties is of serious concern in 21-cm experiments. If the RFI is bright enough, it is mostly flagged by the RFI detection algorithms. On the other hand, the impact of residual transient and non-stationary RFI that stays below the thermal noise level and remains undetected by traditional RFI detection methods is less understood. First, we investigate whether the transient RFI that we observed over different nights correlate with the LST, and if so, at what level? This is a vital test because any contamination that correlates over time is expected to impact the power spectrum and it needs to be mitigated to avoid any biases in the final deep integration power spectra and their interpretation. We use the cross-coherence metric in Fourier space to study the temporal correlation of the transient RFI. For a given pair of gridded visibility cubes $\Tilde{T}_i(\vect{k})$ and $\Tilde{T}_j(\vect{k})$ in Fourier space, we define the 2D cross-coherence, $C_{ij}(k_\perp,k_\parallel)$, as:
\begin{equation}\label{eqn:cross-coherence}
    C_{ij}(k_\perp,k_\parallel) = \dfrac{\langle |\Tilde{T}_{i}^{*}(\vect{k}) \Tilde{T}_{j}(\vect{k})| \rangle^2}{\langle | \Tilde{T}_{i}(\vect{k})|^2\rangle \langle |\Tilde{T}_{j}(\vect{k})|^2\rangle }.
\end{equation}

The $C_{ij}(k_\perp,k_\parallel)$ can have a value between $0-1$ that corresponds to no correlation (zero) or maximum correlation (one) because it is a normalised quantity. This definition is similar to the one used by \cite{mertens2020}. We used only the real component of the cross-power spectrum to calculate the cross-coherence instead of using the complex values. The averaging was performed over all $(k_\perp,k_\parallel)$ modes in the range $0.14\lesssim k\lesssim2.83$ and a single $C_{ij}$ value was obtained for each pair of image cubes. We used all 1-minute image cubes for all 8 nights and calculated their cross-coherences. We trimmed the image cubes to an FoV angle diameter of $120^{\circ}$ (or $\pi$~sr\footnote{We calculate the solid angle as $\Omega = 2\pi (1-\cos{\theta/2})$, where $\theta$ is the angle diameter.} in terms of solid angle footprint), and applied a spatial Hann taper \citep{blackman1958} before Fourier transforming the images, to avoid stationary RFI on the ground and horizon (which is easier to mitigate and not a topic of this study on transient sky-based RFI). We note that the Fourier Transform of full-sky (curved) images under the flat-sky approximation can lead to geometric effects, but we since perform a relative comparison of nights with and without noise, this assumption does not affect our findings and conclusions. In forthcoming analyses, we plan to investigate the application of spherical harmonics-based inversion methods \citep{ghosh2018} for the ACE 21-cm power spectrum measurement.

\begin{figure}
    \centering
   \includegraphics[width=\columnwidth]{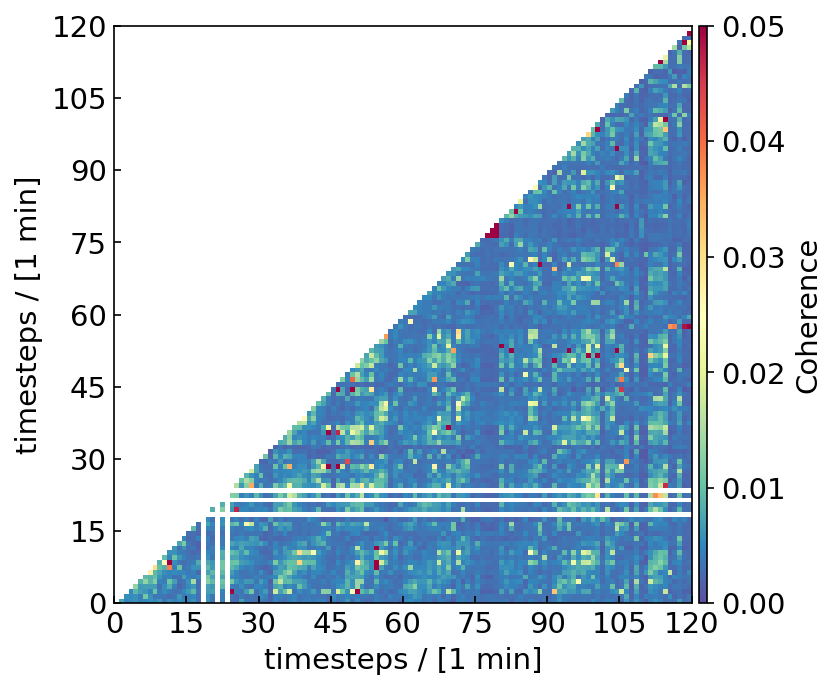}
    \caption{Average cross-coherence of all image-cube pairs in the absence of thermal noise. The grid lines separate timesteps corresponding to different nights. The blank pixels (between 15-30) belong to timesteps that do not have any detected RFI.}
    \label{fig:coherence_no_noise}
\end{figure}

Figure~\ref{fig:coherence_no_noise} shows the average cross-coherence of transient RFI in the absence of noise. We find that for most image-cube pairs the coherence is below 0.01 and only a tiny fraction of pairs, about $7.5$~per cent show coherence higher than 0.01 (and about $1.7$~per cent of pairs show coherence greater than 0.02). The image cube pairs with high coherence and short temporal scale are likely to be associated with transient RFI that is largely incoherent over time. However, we do observe slightly increased coherence (albeit lower than 0.02) of 1-minute image cube pairs corresponding to the same LST. This manifests itself as diagonal structures in figure~\ref{fig:coherence_no_noise} in some pairs of nights. We suspect that this behaviour could be due to some low-level correlated foreground (and their sidelobes) and noise components that remain in the filtered images (overlapping with the pixels with transient RFI) being treated as transient RFI. 

Next, we introduced three different thermal noise levels to the gridded visibilities of every 1~minute image cube to assess the correlation of transient RFI in the presence of thermal noise. The thermal noise rms per gridded visibility cell ($\sigma_{\text{grid}}$) can be calculated as \citep{thompson2001}:
\begin{equation}
    \sigma_{\text{grid}} = \dfrac{\text{SEFD}}{N_\text{vis}\sqrt{2\Delta \nu \Delta t}},
\end{equation}
where SEFD is the system equivalent flux density, $N_\text{vis}$ is the number of visibilities in a $uv$-cell, $\Delta \nu=65$~kHz is the channel width, $\Delta t=4$~seconds is the integration time of a single visibility. To obtain the thermal noise level for deep integrations, we scaled down $\sigma_\text{grid}$ by multiplying it by the factor $(t_\text{image}/t_\text{obs})^{1/2}$, where $t_\text{image}=1$~minute (observing time of a single data-cube), and $t_\text{obs}$ is the total observing time per data-cube for deep integration. We used three different observing times to calculate the thermal noise, $t_{\text{obs}} = 1~$minute, $4.17$~hours, and $25$~hours per data cube which is equivalent to the total deep integration ($t_\text{int}$) of $2~$hours, $500~$hours, and $3000~$hours, respectively, for the full dataset containing 120 such data cubes. Independent noise realisations were generated for each data cube and integration time and added to the corresponding gridded visibilities. For the first two cases, we chose the SEFD of 1.8~MJy which is approximately equal to the observed sensitivity of AARTFAAC-12 at 74~MHz. However, for the third case, we chose a hypothetical instrument twice as sensitive as AARTFAAC-12 and a SEFD of 900~kJy\footnote{Assuming $T_{\text{sys}}\approx T_{\text{sky}} = 50\lambda^{2.56}$~K, this is the minimum achievable SEFD for a sky temperature dominated dipole receiver with the effective area ($A_\text{eff}\approx 5.33$~m$^2$) same as of a LOFAR-LBA dipole at 75~MHz.}. Table~\ref{tab:noise_details} summarises the parameters used to calculate the thermal noise.


\begin{table}
	\centering
	\caption{Thermal noise calculation parameters}
	\label{tab:noise_details}
	\begin{tabular}{lcc} 
		\hline		
		\textbf{SEFD} & \multicolumn{2}{c}{\textbf{Integration time}} \\ 
		 & Single data-cube ($t_\text{obs}$) & Full dataset ($120\times t_\text{obs}$)\\	
		\hline
		1.8~MJy & 1~minute & 2~hours \\ 
		1.8~MJy & 4.17~hours & 500~hours \\ 
		900~kJy & 25~hours & 3000~hours \\ 
		\hline
	\end{tabular}
\end{table}

\begin{figure*}
    \centering
   \includegraphics[width=\textwidth]{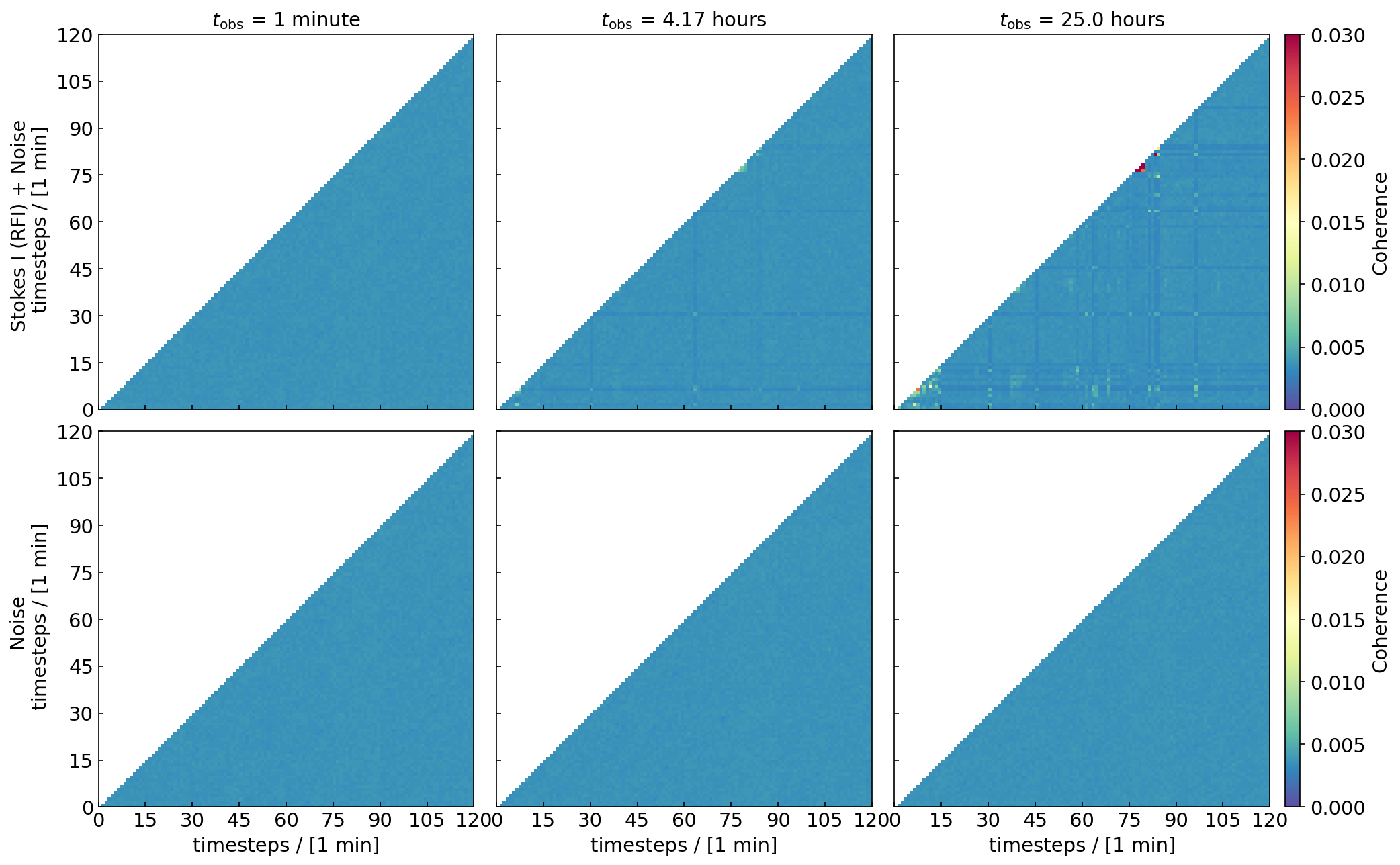}
    \caption{Average cross-coherence of all image-cube pairs. The top row shows cross-coherence of transient RFI in Stokes I images in the presence of thermal noise with different levels (left to right columns: $t_\text{obs}=1$~minute, $4.17$~hours, $25$~hours), and the bottom row shows the cross-coherence of the corresponding thermal noise levels. The transient RFI coherence in the presence of thermal noise is visually indiscernible from the noise coherence itself.}
    \label{fig:rfi_cross_coh_wnoise}
\end{figure*}

 Figure~\ref{fig:rfi_cross_coh_wnoise} shows the cross coherence in the presence of the thermal noise with different levels as described earlier. We observe that the average cross-coherence of transient RFI in the presence of thermal noise virtually disappears for the realistic noise level (1-minute integration time for a 1-minute datacube) and it is visually identical to the cross-coherence of the thermal noise itself. We start to observe some coherence between a small fraction of image cube pairs for the low thermal noise scenarios. We note that the low-level structures that we observe in Fig.~\ref{fig:coherence_no_noise} disappear in the presence of thermal noise because the unmasked region is a tiny fraction of the sky and the low-level correlated component in the unmasked region is subdominant in the cross-coherence compared to the simulated noise.
 
\begin{figure*}
    \centering
   \includegraphics[width=\textwidth]{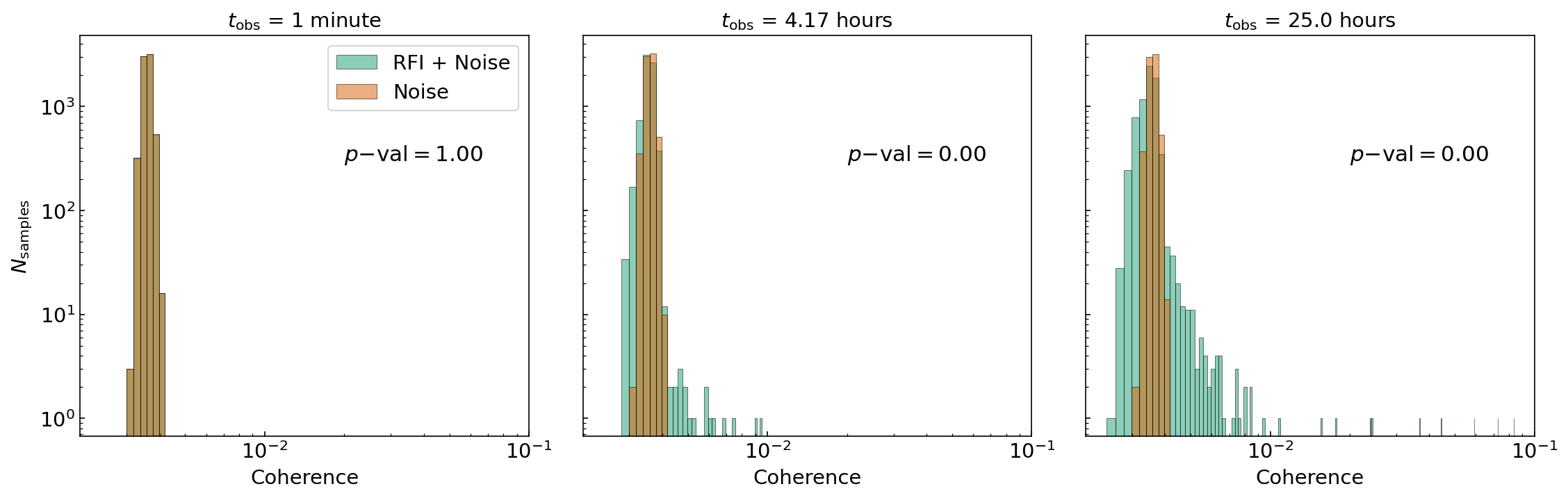}
    \caption{Distribution of average cross-coherences of transient RFI in the presence of thermal noise (shown in green) and cross-coherences of thermal noise (shown in orange). The left, middle, and right panels correspond to different integration times per image-cube, $t_\text{obs} =$ 1~minute, 4.17~hours, and 25~hours, respectively.}
    \label{fig:distribution_ks}
\end{figure*}

 We also perform a two-sample Kolmogorov-Smirnov test to quantify how the transient RFI coherence distribution, in the presence of thermal noise, compares with the coherence distribution of the thermal noise. Figure~\ref{fig:distribution_ks} shows the coherence distributions of the transient RFI in the presence of thermal noise and the thermal noise itself. The $p$-values for the three thermal noise scenarios are $1.0$, $0.0$, and $0.0$, respectively. This suggests that in the realistic thermal noise scenario with an integration time of 1~minute per image-cube, transient RFI coherence is indiscernible from the thermal noise coherence for the AARTFAAC-12 instrument. For the integration time of 4.17~hours per image cube (equivalent to the current total integration time in hand per 1 out of 120 image cubes), the transient RFI coherence distribution becomes slightly skewed, where a tiny fraction of image-cube pairs show slightly higher coherence but still lower than 0.01. However, for a hypothetical instrument twice as sensitive as AARTFAAC-12, the transient RFI starts to appear above the thermal noise for 25 hours of integration per image cube. Still, the transient RFI observed is slow enough that it is detectable (directly or via the PSF sidelobes) by integrating over 1-minute time scales. Therefore, the sky-based masking could also be used to filter this transient RFI from the gridded visibilities before foreground removal and power spectrum estimation using Gaussian Process Regression  \citep{mertens2018,mertens2023} to avoid any low-level biases in the 21-cm signal measurement with extremely sensitive instruments. However, such a strategy to mitigate the transient RFI will be investigated in detail in future work.

\subsection{Effect of a spatial taper}\label{subsec:spatial_taper}
The next part of the study is to investigate the effect of a spatial taper and trimming in the image domain on the transient RFI power in Fourier space. For this test, we use two types of spatial tapering functions, a Hann window and a Tukey or a tapered cosine window \citep{harris1978} with taper parameter $\alpha=0.2$ which means 80~per cent inner region of the image is covered by flat-top of the window and the remaining 20~per cent region is cosine tapered. Applying a Hann window reduces the effective fraction of the image to 25~per cent, whereas the Tukey window with $\alpha=0.2$ reduces the effective fraction to 81~per cent of the image. We also use three different FoV trims, full-sky ($2\pi$~sr), $150^\circ$ ($1.48\pi$~sr) , and $120^\circ$ ($\pi$~sr)  diameter (same as in \citetalias{gehlot2020}). We use the cross-coherence metric (Eq.~\ref{eqn:cross-coherence}) to quantify the impact of the taper and trimming of images on the behaviour of transient RFI in Fourier space. 

\begin{figure*}
    \centering
   \includegraphics[width=\textwidth]{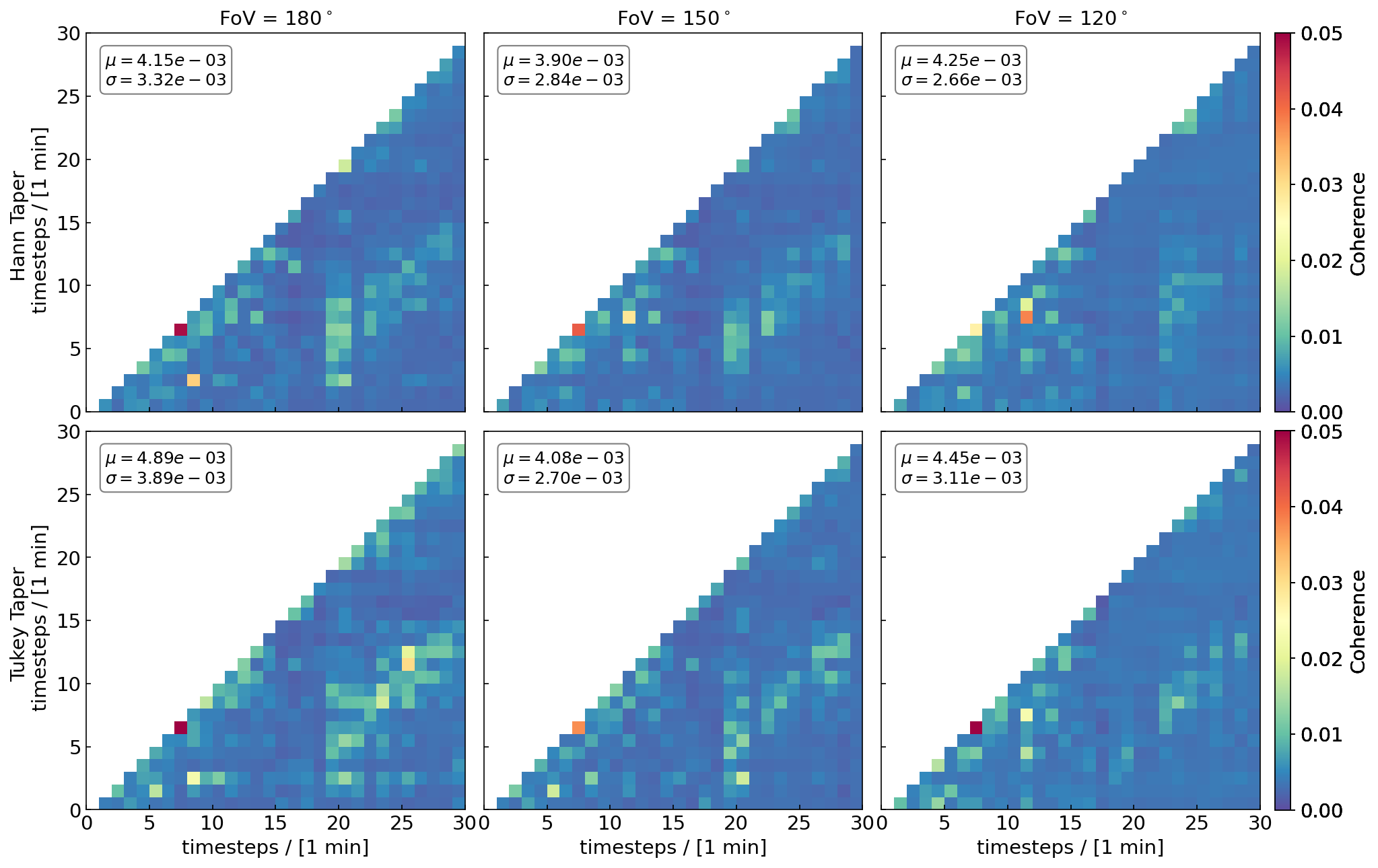}
    \caption{Cross-coherence for all image-cube pairs for the two nights described in Sect.~\ref{subsec:spatial_taper}. The top and bottom rows correspond to Hann and Tukey taper, respectively. Columns from left to right correspond to FoVs of $180^{\circ}$ (no trimming), $150^{\circ}$, and $120^{\circ}$ (same as in \citetalias{gehlot2020}), respectively. The dotted horizontal and vertical lines separate the two nights. The legend in each panel shows the mean and standard deviation of the coherences.}
    \label{fig:coherence_taper}
\end{figure*}

We select two nights that show significant transient RFI levels, that is, nights corresponding to the first and third panels of the top row in Fig.~\ref{fig:RFI_per_night} and calculate cross-coherence for all combinations of 1~minute image-cubes from the two nights. These two nights represent the worst-case scenarios in the set of nights we have used for this work. Figure~\ref{fig:coherence_taper} shows the transient RFI cross-coherence for Hann (top row) and Tukey tapers (bottom row) with different image trimming levels (different columns). We observe that the Tukey taper with $\alpha=0.2$ has an overall worse performance compared to the Hann window regardless of the trimming used. This is mainly due to the decreased sensitivity towards lower elevation caused by the Hann window but with a trade-off of sky sensitivity compared to a Tukey ($\alpha=0.2$) taper. For both Hann and Tukey windows, decreasing the FoV leads to a decrease in overall coherence. This is expected as excluding the regions near the horizon showing high transient RFI reduces the overall contamination in the field. A low-level leakage is still expected due to side lobes. However, it does not correlate over time because the transient RFI is expected to be uncorrelated. Additionally, the average coherences for the different cases are similar ($\sim 4\times10^{-3}$). Whereas, the standard deviation of the coherence values decreases as the FoV is decreased and is slightly lower for the Hann taper compared to that of the Tukey taper. From the findings above, we conclude that using a Hann taper and a FoV of $120^{\circ}$ is a better choice for ACE analysis compared to the Tukey ($\alpha=0.2$) taper to reduce the contamination from transient RFI at low elevations. This was our choice of the taper used for the analysis presented in \citetalias{gehlot2020}. 


\begin{figure*}
    \centering
   \includegraphics[width=\textwidth]{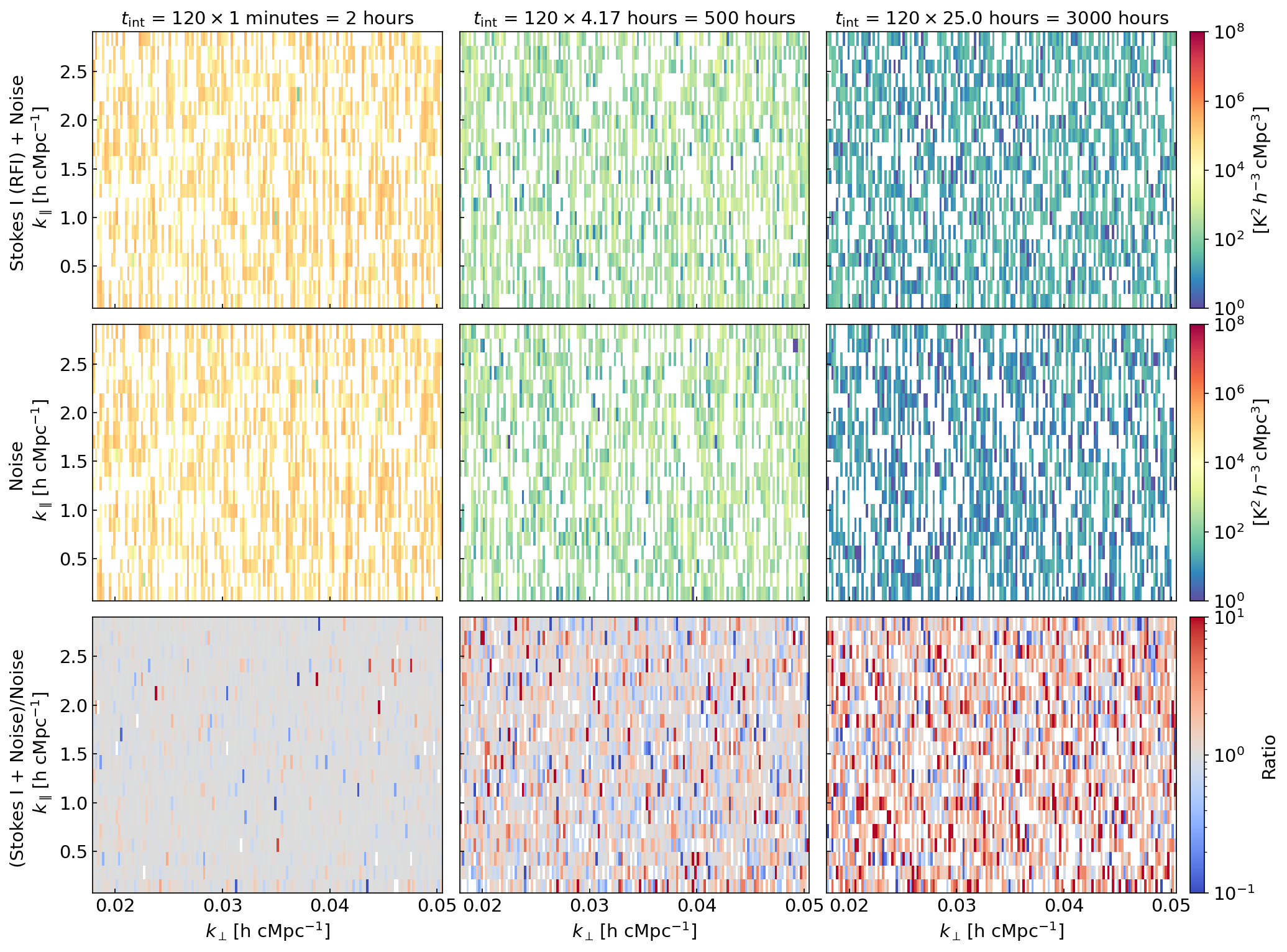}
    \caption{Inverse variance weighted average of the cylindrical cross-power spectra of all image-cube pairs. The top row shows the power spectra of transient RFI in the presence of thermal noise for different deep integration times (left to right columns), the middle row shows the cross-power spectra of the thermal noise itself, and the bottom row shows the ratio of the top row and the middle row. The blank pixels correspond to negative power which occurs due to the presence of the uncorrelated nature of data.}
    \label{fig:2dpspec_rfi}
\end{figure*}

\begin{figure*}
    \centering
   \includegraphics[width=\textwidth]{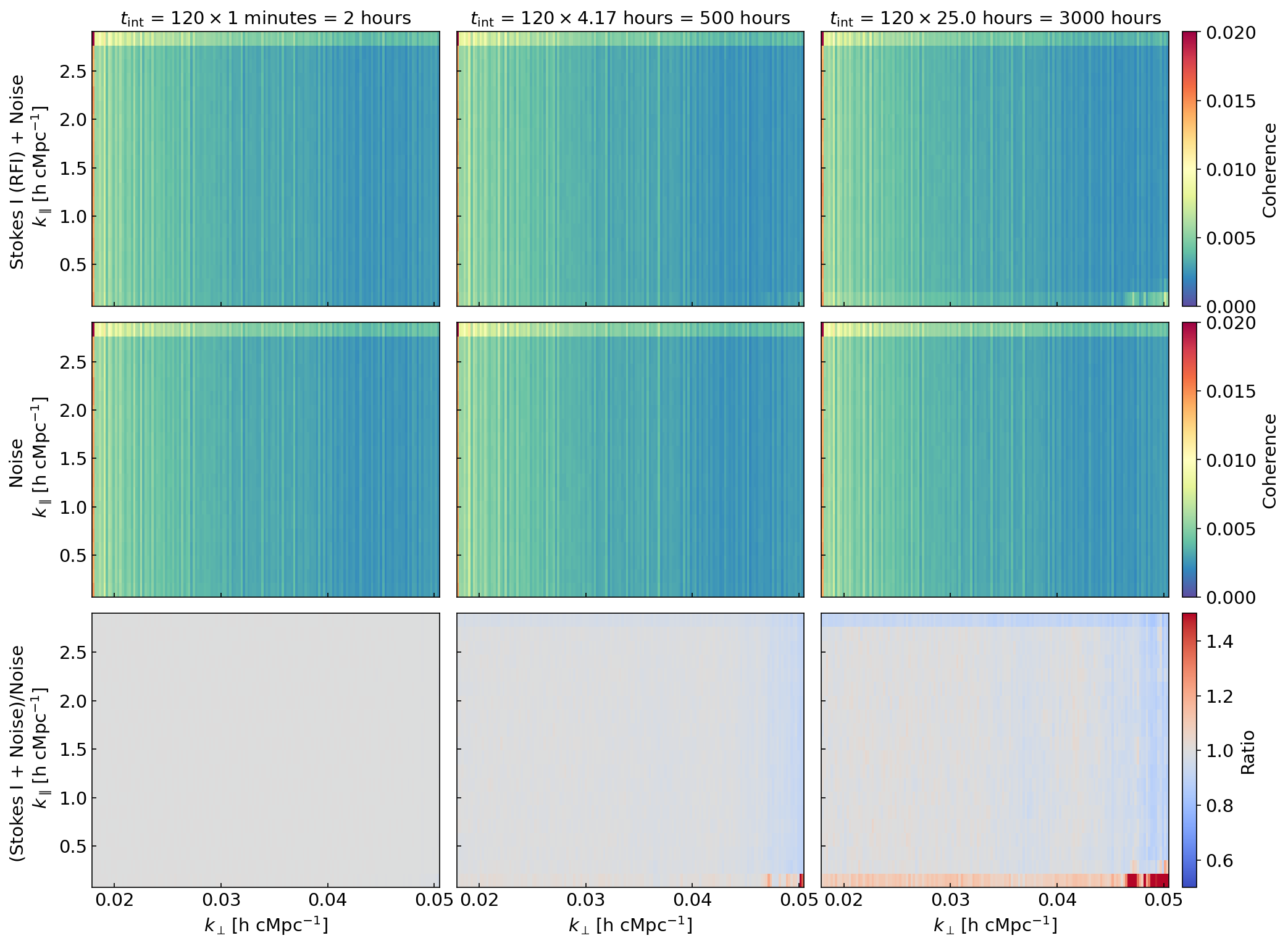}
    \caption{Average cross-coherence of all image-cube pairs. The top row shows the cross-coherence of the transient RFI in the presence of thermal noise for different deep integration times (left to right columns), the middle row shows the cross-coherence of the thermal noise itself, and the bottom row shows the ratio of the top row and the middle row.}
    \label{fig:2dcoherence_rfi}
\end{figure*}

\subsection{Power spectra of transient RFI}

In 21-cm cosmology, cylindrically and spherically averaged power spectra are the most preferred and widely used statistics for diagnostic purposes, investigating the impact of various contaminations, as well as the measurement of the redshifted 21-cm signal from the CD and EoR. Therefore, it is useful to assess the impact of the transient RFI on these two power spectra. We use the standard definition of the cosmological power spectra \citep{morales2004,morales2019} to compute cylindrically and spherically averaged power spectra for each transient RFI image cube in the dataset. However, we choose to calculate cross-power spectra by cross-multiplying two different gridded visibility cubes, instead of calculating the auto-power spectrum by multiplying a gridded visibility cube by itself, to avoid the bias due to the noise and incoherent RFI. This allows us to inspect the correlation of transient RFI. We then average all cross-power spectrum pairs using inverse variance weighting to obtain a single cylindrical and spherical power spectrum. For this test, we use gridded visibilities from the transient RFI image cubes with thermal noise (the three scenarios used in Sect.~\ref{subsec:rfi_correlation_time}) and corresponding thermal noise realisations to calculate cylindrically and spherically averaged power spectrum. In addition to cross-power spectra, we also compute the cylindrically averaged cross-coherence which is essentially a normalised cross-power spectrum. This is done using Eq.~(\ref{eqn:cross-coherence}), but cylindrical averaging is performed instead of averaging all $k$-modes together.

\begin{figure}
    \centering
   \includegraphics[width=\columnwidth]{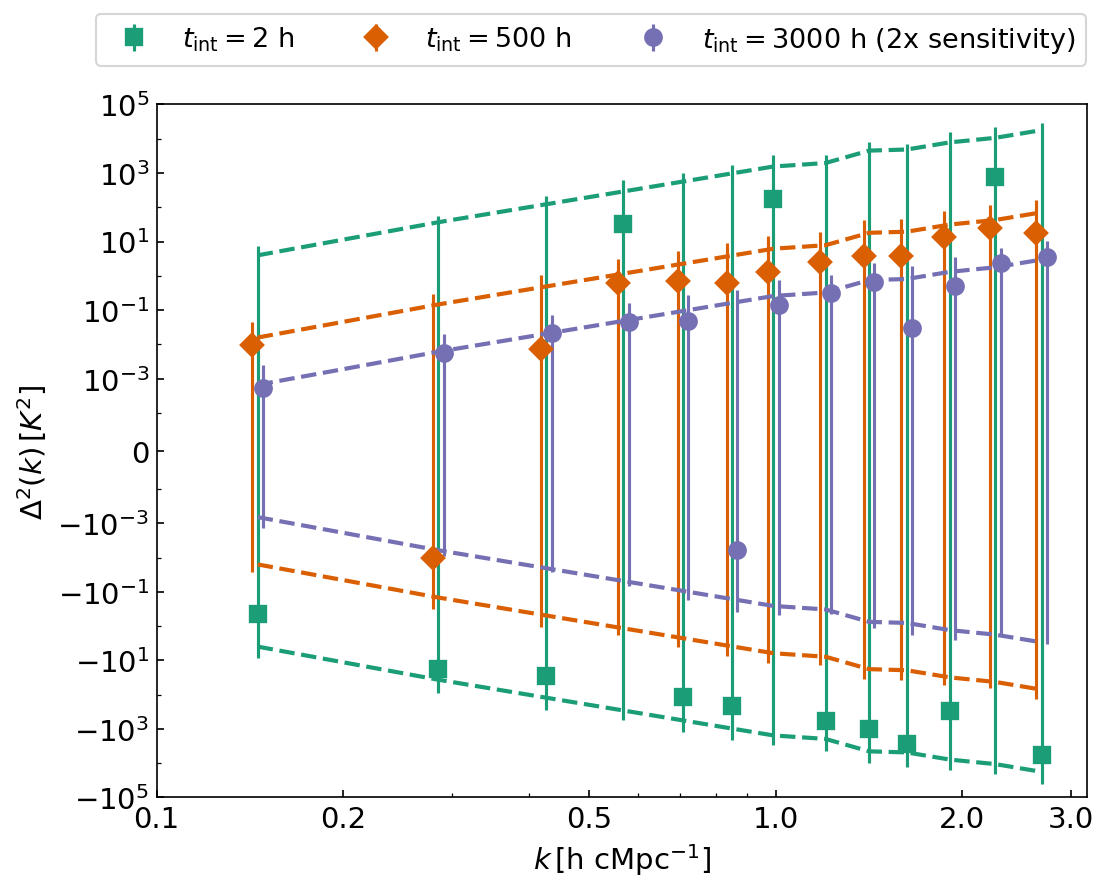}
    \caption{Inverse variance weighted spherically averaged power spectrum of transient RFI in the presence of thermal noise. Different markers (and colours) show transient RFI cross-power spectra for different thermal noise levels and the dashed lines correspond to the $1\sigma$ error due to the thermal noise. We note that the $k$-values for 500~hours and 3000~hours have been shifted slightly to better visualise the error bars. The dashed lines also correspond to the power spectrum sensitivity of the given integration time and instrument sensitivity.}
    \label{fig:3dpspec_rfi}
\end{figure} 
 
Figure~\ref{fig:2dpspec_rfi} shows the inverse variance weighted average of the cross-power spectra calculated from all image-cube pairs corresponding to the eight nights. We do not include auto-power spectra in the averaging to avoid incoherent RFI and noise bias in the final spectrum. We observe that the cylindrically averaged power spectra for transient RFI behave like uncorrelated noise regardless of the integration time. Although it is not very straightforward to interpret the power spectra with higher integration time, we can think of them as if a similar RFI level is observed over the given integration time or the instrument has higher sensitivity. This suggests that the transient RFI we observe is uncorrelated when different nights are combined and leaves little to no impact on $k$-modes where we expect to measure the faint 21-cm signal from Cosmic Dawn. The ratio of the power spectra of transient RFI and the thermal noise has a median of about unity for the three scenarios. However, the variance of the ratio increases as the thermal noise decreases. This is likely due to the presence of uncorrelated transient RFI that enhances the variance relative to the thermal noise. For the integration time of 500~hours, we do not observe a significant difference between the transient RFI power and the thermal noise power spectrum. This is an important result because RFI is considered one of the major hurdles in 21-cm cosmology experiments. However, this conclusion should be treated with some care because it only applies to the transient RFI that is sky-based, varies over time, is above the location of the LOFAR core, is in the frequency range studied here, and does not show any night-to-night correlation at the same LST. Persistent RFI (whether narrow band or broadband) from sources such as geostationary satellites, FM transmissions, digital audio and video broadcasts contaminates the observed data at approximately the same level and location (in frequency spectrum) regardless of the sidereal time of the observation and leaves strongly correlated artefacts in the data. The latter contamination is of grave concern in 21-cm experiments and needs to be carefully mitigated. 

We can draw similar conclusions from Fig.~\ref{fig:2dcoherence_rfi} that shows the cylindrical cross-coherence averaged over all image-cube pairs. We do not observe any discernible structure due to the transient RFI in addition to the thermal noise at different $k$-modes for 2~hours and 500~hours of integration, suggesting that the transient RFI does not correlate over time and combining observations of transient RFI localised differently in image space does not result in correlated structures in Fourier space. We start to see very low-level coherence at the smallest $k_{\parallel}$ mode for the third scenario which we suspect is due to spectrally smooth foregrounds (and sidelobes) that are falsely detected as transient RFI. Finally, we show inverse variance weighted averages of the spherically averaged cross-power spectra for all image-cube pairs in Fig.~\ref{fig:3dpspec_rfi}. We see that for all three scenarios, 2~hours, 500~hours, and 3000~hours (and twice the AARTFAAC-12 sensitivity) integration, the transient RFI power in the presence of thermal noise is consistent with the thermal noise floor within $2\sigma$ error.

\section{Conclusions and discussion}

The LOFAR-AARTFAAC Cosmic Explorer project aims to measure the power spectrum of the redshifted 21-cm signal of neutral hydrogen from the Cosmic Dawn ($z\sim18$) using the AARTFAAC-12 wide-field imager of LOFAR-LBA and to constrain or exclude non-standard (``exotic'') astrophysical models that could explain the unusual nature of the putative 21-cm absorption feature as reported by the EDGES collaboration \citep{bowman2018}. In \citetalias{gehlot2020}, we presented the first upper limits on the redshifted 21-cm signal at $z=18.26$ using a small subset of AARTFAAC data recorded as a part of the ACE observing campaign. However, since the LOFAR low-band receivers that compose AARTFAAC have a full-sky FoV and are situated in a high population density area of Western Europe, they may be especially susceptible to the radio frequency interference at low elevations as well as sources of interference that rapidly move across the FoV, for example, aeroplanes,  satellites, meteors etc. In this work, we investigated the impact of transient RFI on the power spectrum which is the main statistic that we use to measure the redshifted 21-cm signal. Our findings are summarised below:

\begin{enumerate}
    \item Different levels of transient RFI are found for different nights. Most of the RFI shows up closer to the horizon, but we also observe rapidly moving RFI sources that cross over the FoV of AARTFAAC within a few minutes. Most of the transient RFI appears to be narrow-band and localised in image space. They are most likely aeroplane communication beacons in the observed spectral window.
    
    \item A cross-coherence analysis shows that the transient RFI between different points in time (image-cubes) exhibit little to no correlation ($2$~per cent or below) except for a few outliers that show a low-level correlation of the order of a few per cent. Additionally, in the presence of thermal noise, the RFI coherence behaviour is mostly indistinguishable from the thermal noise, except for the extremely low noise regime (equivalent to $>500$~hours of integration) where outliers start to appear.
    
    \item Trimming the images to a certain FoV or applying a spatial taper to exclude the transient RFI at low elevation has a substantial impact on the coherence of the transient RFI in Fourier space. For example, trimming the image to a smaller FoV of $120^\circ$ reduces the RFI coherence compared to the full FoV. Similarly, applying a Hann spatial taper shows relatively lower transient RFI coherence compared to a Tukey taper, suggesting that most RFI is generated on the ground plane and/or near the horizon.
    
    \item The cylindrically averaged power spectrum of the transient RFI in the presence of thermal noise and after applying a Hann spatial taper behaves similarly to uncorrelated noise regardless of the different thermal noise levels. However, the overall variance of the transient RFI relative to the thermal noise increases for deep integration of 500~hours and above. Similarly, cylindrically averaged cross-coherence of transient RFI shows no additional coherence in Fourier space for up to at least 500~hours of integration. However, the 3000~hours of integration with the hypothetical instrument twice the sensitivity of AARTFAAC shows very low-level coherence at the smallest $k_{\parallel}$ mode which is likely due to residual correlated foregrounds remaining after the filtering process. The spherically averaged power spectrum of transient RFI in the presence of thermal noise is consistent with instrument sensitivity for all integration times, 2~hours, 500~hours, as well as $3000$~hours for a hypothetical instrument twice as sensitive as AARTFAAC.
\end{enumerate}


From the investigation presented in this work, we conclude that the sky-based transient RFI at the location of LOFAR core (located in Exloo, Netherlands) between 72-75~MHz for any practical purpose is uncorrelated over time, and indistinguishable from the thermal noise equivalent to 500~hours of integration, which is the same as the amount of data recorded during the ACE observing campaign. Additionally, following a more conservative approach in future, we plan to investigate advanced mitigation techniques for transient RFI such as:

\begin{enumerate}
    \item \textit{Image-based filtering of transient RFI}: In the analysis, we observed that the transient RFI is localised in image space over short periods. A 4D mask can be used to filter out the voxels (in space, frequency and time) affected by transient RFI. These voxels can then be filled by artificial emission derived from a physically and statistically motivated model of the structure (foregrounds and noise) using a robust approach such as Gaussian Process Regression \citep{mertens2018}, to avoid artefacts in the power spectrum. However, an in-depth investigation of the impact of such an approach on 21-cm signal measurement is still required.
    
    \item \textit{Flagging of RFI in visibility space}: Visibilities observed with AARTFAAC have a high noise level because of the small effective collecting area of receiver elements. Traditional algorithms that operate on per baseline dynamic spectra of visibilities to search for outliers, therefore, are sub-optimal for detecting low-level transient RFI. The use of baseline-integrated statistics provides a method to search for faint RFI signatures from aeroplanes and digital broadcasts. \cite{offringa2015} shows an example of applying their detection method on the baseline-detected standard deviation, thereby detecting RFI that is not detectable in a single baseline. Similarly, \cite{wilensky2019} performs detection on the incoherently averaging the visibility spectra of the baselines to improve detection. This method is shown to be particularly effective for MWA datasets. However, it may lead to significant over-flagging because of the averaging of multiple baselines, which are all flagged even if only a subset exhibits RFI. We plan to explore a similar approach by grouping baselines based on length and orientation to avoid over-flagging. Additionally, we plan to explore the use of cross-polarisations for faint RFI detection  \citep{yatawatta2020} in future analyses.
    
\end{enumerate}

In closing, significantly deeper observing campaigns with highly sensitive instruments may be susceptible to low-level transient RFI and would require a transient RFI mitigation strategy. However, conclusions from this investigation can be extrapolated to the next generation of 21-cm cosmology experiments such as with the SKA and HERA, which are located at more radio-quiet sites and have a narrower FoV. These experiments are therefore less likely to suffer from contamination from the transient RFI that is uncorrelated over time impacting the redshifted 21-cm signal measurement.

\begin{acknowledgements}
BKG, LVEK, SAB, SG, CH, and SM acknowledge the financial support from the European Research Council (ERC) under the European Union’s Horizon 2020 research and innovation programme (Grant agreement No. 884760, "CoDEX"). LVEK, ARO, and EC acknowledge support from the Centre for Data Science and Systems Complexity (DSSC), Faculty of Science and Engineering at the University of Groningen. FGM acknowledges support from a PSL Fellowship. This work was supported in part by ERC grant 247295 ``AARTFAAC" to RAMJW. LOFAR, the Low-Frequency Array designed and constructed by ASTRON, has facilities in several countries, owned by various parties (each with their own funding sources), and collectively operated by the International LOFAR Telescope (ILT) foundation under a joint scientific policy. 
This research made use of publicly available software developed for LOFAR and AARTFAAC telescopes. Listed below are software packages used in the analysis: \\
\textsc{aartfaac2ms} (\url{https://github.com/aroffringa/aartfaac2ms}),\\ 
\textsc{aoflagger} (\url{https://gitlab.com/aroffringa/aoflagger}), \\
\textsc{dp3} (\url{https://git.astron.nl/RD/DP3}), \\
\textsc{wsclean} (\url{https://gitlab.com/aroffringa/wsclean}).\\ 
\textsc{ps\_eor} (\url{https://gitlab.com/flomertens/ps_eor}).\\ 
\\
The analysis also relies on the Python programming language (\url{https://www.python.org}) and several publicly available python software modules: \\
\textsc{astropy} (\url{https://www.astropy.org/}; \citealt{astropy2013}), \\
\textsc{matplotlib} (\url{https://matplotlib.org/}),\\ 
\textsc{scipy} (\url{https://www.scipy.org/}), and \\
\textsc{numpy} (\url{https://numpy.org/}).
\end{acknowledgements}

\bibliographystyle{aa}
\bibliography{rfi_analysis}

\end{document}